\documentstyle[12pt,fleqn]{article}
\topmargin - 1cm

\catcode`@=11
\@addtoreset{equation}{section}
\catcode`@=12
\begin{document}
\title{\vskip-1.7cm \bf  Quantum Dirac constraints, Ward identities and path
integral in relativistic gauge}
\date{}
\author{A.O.Barvinsky}
\maketitle
\hspace{-8mm}{\em
Theory Department, Lebedev Physics Institute and Lebedev Research Center in
Physics, Leninsky Prospect 53,
Moscow 117924, Russia}\footnote
{Electronic address: barvin@td.lpi.ac.ru}
\begin{abstract}
Quantum Dirac constraints in generic constrained system are
solved by directly calculating in the one-loop
approximation the path integral with relativistic
gauge fixing procedure. The calculations
are based on the reduction algorithms for functional determinants
extended to gauge theories. Explicit mechanism of transition from
relativistic gauge conditions to unitary gauges, participating
in the construction of this solution, is revealed by the method of
Ward identities.
\end{abstract}

\section{Introduction}
\hspace{\parindent}
One of the old problems in the present day high energy physics consists in
the development of path integration in Dirac quantization of
constrained systems -- the only known regular method in problems that
go beyond the scope of exactly solvable models. In contrast with non-gauge
theories, the Schrodinger equation in such systems is supplemented by
quantum Dirac constraints on quantum states. Moreover, in parametrized
systems with a vanishing Hamiltonian there is no independent
Schrodinger equation and quantum dynamics is encoded in the Dirac
constraints along with their gauge invariance properties. For such
problems the fundamental dynamical equations are not of a manifestly
evolution type and, therefore, the path integral applications are
much less straightforward.

One of the first formulations of the path integral as a solution of
the non-evolutionary quantum Dirac constraints belongs to H.Leutwyler
\cite{Leutw} who proposed a special solution of the Wheeler-DeWitt
equations in the form of a naive functional integral (lacking the correct
gauge-fixing procedure). The path integral in unitary
gauge with exhaustive set of boundary conditions was
later proposed in \cite{BAO}. Then this canonical path integral
was converted to the spacetime covariant form of the functional integral
over Lagrangian variables and ghost fields in relativistic gauge
\cite{HH,barvin,HalHar,BarvU}.

Of course, these results encorporate a well-known statement on
equivalence of the canonical and covariant quantizations pioneered in
\cite{FV}. In contrast with this, the focus of \cite{BAO,HH,barvin,HalHar}
was on nontrivial boundary conditions in spacetime. Correct treatment
of these boundary conditions
leads to the proof that this path integral solves quantum Dirac
constraints (\ref{0.0}). However, this proof \cite{BAO,HH,barvin0,HalHar}
has a formal nature and does not even allow one to fix
the operators of these constraints which in a rather uncontrollable
way depend on the calculational method for a path integral \cite{barvin0}.
Thus, no check of the solution to quantum constraints was thus far given
by direct calculations of the path integral. The goal of this paper is to
perform such a check for generic systems subject to first class constraints
in the one-loop approximation of semiclassical expansion.

The calculations are based on the reduction methods for one-loop
functional determinants \cite{reduct} extended here to the case of special
boundary conditions characteristic of the coordinate representation in
Dirac quantization method. These calculations lead to the Van Vleck-type
solution of quantum Dirac constraints obtained earlier in the series of
author's works \cite{GenSem,BKr,BarvU} and as a byproduct establish the
explicit mechanism of transition from relativistic gauges in the
path integral to unitary gauges participating in the canonical construction
of this solution.

\section{Dirac quantization of constrained systems}
\hspace{\parindent}
Consider a theory with the action $S\,[\,g\,]$ invariant under local gauge
transformations of fields $g=g^a$ with generators $R^a_\mu$
	\begin{eqnarray}
	&&R^a_\mu\,\frac{\delta S\,[\,g\,]}
	{\delta g^a}=0,                          \label{1.10}\\
	&&S\,[\,g\,]=
	\int_{t_-}^{t_+}dt\,L(q,\dot{q},N).         \label{1.2}
	\end{eqnarray}
In the form (\ref{1.2}), explicitly featuring
time parameter, gauge fields are decomposed in two sets
$g^a=(q^i(t),N^\mu(t))$ such that the Lagrangian does not involve
time derivatives of $N=N^\mu(t)$ and generates the canonical action
	\begin{eqnarray}
	S\,[\,q,p,N\,]=\int_{t_-}^{t_+}dt\,
	[\,p_i \dot{q}^i-N^\mu T_\mu(q,p)\,]         \label{00.1}
	\end{eqnarray}
with canonical variables $(q,p)=(q^i,p_i)$
and Lagrange multiplyers $N=N^\mu$ for the Dirac constraints
$T_\mu(q,p)$\footnote
{
In quantum mechanical problems the range of indices $i=1,...n$ and
$\mu=1,...m$ is finite and extends to infinity in field models for which
we use condensed notations of two types -- canonical and covariant.
{\it Canonical} condensed indices include spatial labels and their
contraction implies spatial integration. {\it Covariant} condensed
indices involve time, and their contraction also implies
time integration. They can be easily distinguished by the context in
which they appear. As a rule, we imply that condensed indices are
canonical when the time label is explicitly written. For Einstein theory,
for example, $g^a\sim g_{\alpha\beta}({\bf x},t),\,a=(\alpha\beta,{\bf x},t)$,
is the spacetime metric, $q^i=g_{ab}({\bf x}),\,i=(ab,{\bf x})$, is a
spatial metric and $N^\mu\sim g_{0\alpha}({\bf x}),\,\mu=(\alpha,{\bf x})$
are lapse and shift functions .
}.
For the gauge ivariant action the latter belong
to the first class, i.e. satisfy the algebra
	$\{T_\mu,T_\nu\}=U^{\alpha}_{\mu\nu}T_\alpha$
of Poisson brackets with structure functions $U^{\alpha}_{\mu\nu}$.

Dirac quantization consists in promoting
phase-space variables and constraints to the
operator level $(q,p,T_\mu)\rightarrow (\hat{q},\hat{p},\hat{T}_\mu)$ and
selecting the physical states $|\,{\mbox{\boldmath$\Psi$}}\big>$ in the
representation space of $(\hat{q},\hat{p},\hat{T}_\mu)$ by the equation
	$\hat{T}_\mu|\,{\mbox{\boldmath$\Psi$}}\big>=0$
\cite{BF,BarvU}. Operators $(\hat{q},\hat{p})$ are subject
to canonical commutation relations $[\hat{q}^k,\hat{p}_l]=i\hbar\delta^k_l$
and operators $\hat{T}_\mu$ should satisfy the correspondence principle with
classical constraints and generalize the Poisson brackets
algebra to the commutator level
	$[\hat{T}_\mu,\hat{T}_\nu]=
	i\hbar\hat{U}^{\lambda}_{\mu\nu}\hat{T}_\lambda$
with certain operator structure functions $\hat{U}^{\lambda}_{\mu\nu}$.
This algebra serves as integrability conditions for quantum constraints.
In the coordinate representation,
	$\big<\,q\,|\,{\mbox{\boldmath$\Psi$}}\big>=
	{\mbox{\boldmath$\Psi$}}(q),\,\,\,\,
	p_k=\hbar\partial/i\partial q^k$,
the latter become differential equations on the physical wave function
	$\hat{T}_\mu(q,\hbar\partial/i\partial q)\,
	{\mbox{\boldmath$\Psi$}}(q)=0$.

Note that, without loss of generality, the nonvanishing Hamiltonian $H(q,p)$
is not included in (\ref{00.1}). By extending phase space with
extra canonical pair $(q^0,p_0),\,q^0\equiv t,$ subject to the
constraint $p_0+H(q,p)=0$ any system can be reduced to the case
(\ref{00.1}). At the quantum level this extra Dirac constraint becomes
the Schrodinger equation for the wave function with parametrized
time ${\mbox{\boldmath$\Psi$}}(t,q)={\mbox{\boldmath$\Psi$}}(q^0,q)$.
In this sense the dynamical content of any theory can be encoded in
quantum constraints of the above type.

The path integral arises as a special two-point solution
of these equations ${\mbox{\boldmath$K$}}(q,q')$ -- the analogue of the
two-point evolution operator for the Schrodinger equation \cite{Leutw,BAO}
	\begin{eqnarray}
	\hat{T}_{\mu}(q,\hbar\partial/i\partial q)\,
	{\mbox{\boldmath$K$}}(q,q')=0.                     \label{0.0}
	\end{eqnarray}
This is a path integral over the fields in (space)time domain $t_-<t<t_+$
with the boundary conditions related to the arguments of this kernel
\cite{Leutw}. In view of gauge invariance this integral involves a
typical Faddeev-Popov gauge fixing procedure and integration over the
ghosts $C$ and $\bar{C}$ \cite{BAO,barvin}
	\begin{eqnarray}
	&&{\mbox{\boldmath$K$}}(q_+,q_-)=
	\int Dg\,\mu[\,g\,]\,DC\,D\bar{C}
	\exp\frac i\hbar \left\{S_{\rm gf}[\,g\,]
	+\bar{C}_\mu Q^\mu_\nu C^\nu\right\},         \label{2.1}\\
	&&S_{\rm gf}[\,g\,]=S\,[\,g\,]-
	\frac 12\,\chi^\mu\,c_{\mu\nu}\chi^\nu,      \label{2.2}\\
	&&Q^\mu_\nu=\frac{\delta\chi^\mu}
	{\delta g^a}R^a_\nu,                         \label{2.6}\\
	&&\mu[\,g\,]=\prod_{t}[\,{\rm det}\,G_{ik}(t)\;
	{\rm det}\,c_{\mu\nu}(t)\,]^{1/2}
	\equiv\,\left(\,{\rm Det}\,G_{ik}\;
	{\rm Det}\,c_{\mu\nu}\right)^{1/2}.       \label{2.8}
	\end{eqnarray}
Here the gauge-breaking term in the gauge-fixed action $S_{\rm gf}[\,g\,]$
is quadratic in the {\it relativistic} gauge conditions
	\begin{eqnarray}
	\chi^\mu=\chi^\mu(g,\dot g),\,\,\,
	a^\mu_\nu=-\frac{\partial\chi^\mu}
	{\partial{\dot N}^\nu},\,\,\,
	{\rm det}\,a^\mu_\nu \neq 0,            \label{2.5}
	\end{eqnarray}
with some invertible gauge-fixing matrix $c_{\mu\nu}$,
$Q^\mu_\nu$ is a nondegenerate ghost operator and
$\mu[\,g\,]$ is a local measure with
$G_{ik}=\partial^2 L/\partial\dot q^i\partial\dot q^k$ -- the Hessian
of the classical Lagrangian\footnote
{
The symbol $\rm det$ in (\ref{2.8}) and (\ref{2.5}) denotes
the determinants with respect to {\it canonical} condensed indices in contrast
with the determinants in the space of functions of time denoted by $\rm Det$,
defined with respect to {\it covariant} condensed indices.}.
Integration in (\ref{2.1}) runs over fields with fixed values of canonical
coordinates and ghosts at $t_{\pm}$, while the boundary values of
Lagrange multiplyers are integrated over in the infinite range:
	\begin{eqnarray}
	&&q^i(t_\pm)=q^i_\pm,\,\,\,C^\mu(t_\pm)=0,
	\,\,\,\bar{C}_\nu(t_\pm)=0,                  \label{2.11}\\
	&&-\infty<N^\mu(t_\pm)<+\infty.              \label{2.12}
	\end{eqnarray}

These boundary conditions are invariant with respect to BRST transformations
of the total action in (\ref{2.1}) and enforce the gauge independence
of the path integral \cite{barvin}. They also lead to a formal proof
that this integral solves quantum Dirac constraints in the coordinate
representation of $q=q_+$ and $p=\hbar\partial/i\partial q_+$. This proof
is based on an obvious consequence of the integration range for $N^\mu(t_+)$
	\begin{eqnarray}
	\int Dg\,DC\,D\bar{C}\,
	\frac\delta{\delta N^\mu(t_+)}\,\Big(\,...\,\Big)=0,  \label{2.13}
	\end{eqnarray}
where ellipses denote the full integrand of the path integral (\ref{2.1}).
The functional differentiation here boils down to the
deexponentiation of the constraint
	$\delta S\,[\,g\,]/\delta N^\mu(t_+)=
	-T_\mu(t_+)$
(contributions of gauge-breaking and ghost terms cancel
in virtue of Ward identities \cite{barvin}). This constraint can be
extracted from under the integral sign in (\ref{2.13}) in the form of a
differential operator $\hat{T}_\mu(q_+,\hbar\partial/i\partial q_+)$ acting
on ${\mbox{\boldmath$K$}}(q_+,q_-)$, so that the equation (\ref{2.13}) takes
the form of quantum Dirac constraint (\ref{0.0}) \cite{barvin0}.
This derivation is, however, purely formal and in an
uncontrolable way depends on the skeletonization of the path integral
\cite{barvin0}.

\section{One-loop approximation}
\hspace{\parindent}
Here we compare two semiclassical representations of
${\mbox{\boldmath$K$}}(q,q')$ -- one obtained in \cite{GenSem,BKr} by
directly solving eq.(\ref{0.0}) and another resulting from the one-loop
approximation for the path integral (\ref{2.1}). The
operators $\hat{T}_\mu(q,\hbar\partial/i\partial q)$ closing the commutator
algebra in the first subleading order of $\hbar$-expansion were built in
\cite{BKr}. The solution to the corresponding quantum constraints
(\ref{0.0})
	\begin{eqnarray}
	{\mbox{\boldmath$K$}}(q,q')=
	{\mbox{\boldmath$P$}}(q,q')\,
	e^{\,{}^{\textstyle{\frac i\hbar
	{\mbox{\boldmath$S$}}(q,q')}}}            \label{0.1}
	\end{eqnarray}
is enforced in the same approximation by the following equations for
the Hamilton-Jacobi function ${\mbox{\boldmath$S$}}(q,q')$ and the
preexponential factor ${\mbox{\boldmath$P$}}(q,q')$ \cite{GenSem,BKr,BarvU}
(see also \cite{Kief} in the gravitational context)
	\begin{eqnarray}
	&&T_\mu(q,\partial {\mbox{\boldmath$S$}}/\partial q)=0, \label{0.2}\\
	&&\frac\partial{\partial q^i}
	(\nabla^i_\mu {\mbox{\boldmath$P$}}^2)=
	U^{\lambda}_{\mu\lambda} {\mbox{\boldmath$P$}}^2,  \label{0.3}\\
	&&\nabla^i_\mu\equiv\left.
	\frac{\partial T_\mu}{\partial p_i}
	\right|_{\;\textstyle p=\partial
	{\mbox{\boldmath$S$}}/\partial q}.                    \label{0.4}
	\end{eqnarray}
A particular solution of these equations is generated by the principal
Hamilton function for ${\mbox{\boldmath$S$}}(q,q')$
-- the action calculated at the classical extremal $g=g(\,t\,|q,q')$
which joins the points $q$ and $q'$. This function also satisfies the
Hamilton-Jacobi equations with respect to its second argument
	$T_\mu(q',-\partial {\mbox{\boldmath$S$}}/\partial q')=0$.
The corresponding preexponential factor is a special
generalization of the Pauli-Van Vleck-Morette formula \cite{Morette}
in terms of the Van-Vleck matrix
	\begin{eqnarray}
	{\mbox{\boldmath$S$}}_{ik'}=
	\frac{\partial^2{\mbox{\boldmath$S$}}(q,q')}
	{\partial q^i\, \partial q^{k'}},             \label{0.6}
	\end{eqnarray}
In contrast with non-gauge theories this matrix is degenerate
because it has left and right zero-eigenvalue eigenvectors
\cite{GenSem,BKr,BarvU}
	\begin{eqnarray}
	\nabla^i_\mu{\mbox{\boldmath$S$}}_{ik'}=0,\,\,
	{\mbox{\boldmath$S$}}_{ik'}\nabla^{k'}_\nu=0,\,\,
	\nabla^{k'}_\nu\equiv\left.
	\frac{\partial T_\nu}{\partial p_{k'}}
	\right|_{\;\textstyle p=-\partial
	{\mbox{\boldmath$S$}}/\partial q'}.      \label{0.8}
	\end{eqnarray}
The construction of ${\mbox{\boldmath$P$}}(q,q')$ is, therefore, equivalent
to the gauge-fixing procedure. It consists in adding the ``gauge-breaking''
term bilinear in ``gauge conditions'' $X^\mu_i$ and $X^\nu_{k'}$ -- two
sets of arbitrary covectors at the points $q$ and $q'$
	\begin{eqnarray}
	D_{ik'}={\mbox{\boldmath$S$}}_{ik'}+
	X^\mu_i C_{\mu\nu} X^\nu_{k'},             \label{0.9}
	\end{eqnarray}
which replaces degenerate ${\mbox{\boldmath$S$}}_{ik'}$ by the new
invertible matrix $D_{ik'}$,
provided that the gauge-fixing matrix $C_{\mu\nu}$ is also invertible
and these covectors produce invertible ``Faddeev-Popov operators''
	\begin{eqnarray}
	&&J^\mu_\nu=X^\mu_i\nabla^i_\nu,\,\,\,
	J\equiv{\rm det}\,J^\mu_\nu,             \label{0.10} \\
	&&J'^{\mu}_{\,\nu}=X^\mu_{i'}\nabla^{i'}_\nu,\,\,\,
	J'\equiv{\rm det}\,J'^{\mu}_{\,\nu}.             \label{0.11}
	\end{eqnarray}

Then the solution of the continuity equations (\ref{0.3})
reads \cite{GenSem,BKr,BarvU}
	\begin{eqnarray}
	{\mbox{\boldmath$P$}}(q,q')=
	\left[\frac{{\rm det}\,D_{ik'}}
	{JJ'\,{\rm det}\,C_{\mu\nu}}\right]^{1/2}.      \label{0.12}
	\end{eqnarray}
This solution\footnote
{
Particular example of such solution for a {\it single} Wheeler-DeWitt
equation was found in \cite{Kief}.
}
is a direct analogue of the one-loop effective action in gauge theory
\cite{PhysRep} -- the gauge field contribution ${\rm det}\,D_{ik'}$
partly compensated by the contribution of ghosts $JJ'$.

The covectors $(X^\mu_i,X^\nu_{k'})$ are matrices of gauge
conditions because, as shown in \cite{GenSem,BKr}, the quantum Hamiltonian
reduction of ${\mbox{\boldmath$K$}}(q,q')$ leads to the unitary
evolution operator in the physical sector defined by the {\it unitary}
gauge conditions
	\begin{eqnarray}
	X^\mu(q,t)=0,\,\,\,
	X^\mu_i=\frac{\partial X^\mu}{\partial q^i}.     \label{0.15}
	\end{eqnarray}
Unitary gauge conditions are imposed only on phase space variables
$(q,p)$ (here only on coordinates $q$)\footnote
{
Explicit time dependence of $X^\mu(q,t)$ is necessary in theories
with parametrized time in order to have nontrivial time evolution
with a nonvanishing physical Hamiltonian \cite{BarvU,BKr}.
}.
In contrast with {\it relativistic} gauges involving Lagrange multiplyers,
they manifestly incorporate unitarity but lead to the loss of manifest
covariance.

On the contrary, the path integral (\ref{2.1}) in relativistic gauge is
potentially a spacetime covariant object. Its Feynman diagrammatic
technique was built in \cite{barvin} with
a special emphasis on boundary conditions at $t_\pm$, because in other
respects the $\hbar$-expansion produces a standard set of Feynman graphs.
Thus, the one-loop path integral also has a form
(\ref{0.1}) with the same principal Hamilton function
${\mbox{\boldmath$S$}}(q,q')=S\,[\,g(\,t\,|\,q_+,q_-)\,]$ -- the action
at the solution $g=g(\,t\,|\,q_+,q_-)$ of the following boundary value
problem in the gauge (\ref{2.5}) \cite{barvin}
	\begin{eqnarray}
	&&\frac{\delta S\,[\,g\,]}
	{\delta g^a(t)}=0,\,\,\,\chi^\mu(g,\dot{g})=0,\,\,\,
	t_-\leq t\leq t_+,                         \label{2.16}\\
	&&q(t_\pm)=q_\pm,                          \label{2.17}
	\end{eqnarray}
and the one-loop preexponential factor
	\begin{eqnarray}
	{\mbox{\boldmath$P$}}(q_+,q_-)=
	\left(\frac{{\rm Det}\,F_{ab}}
	{{\rm Det}\,a_{ab}}\right)^{\!-1/2}
	\left.\frac{{\rm Det}\,Q^\mu_\nu}
	{{\rm Det}\,a^\mu_\nu}\,
	\right|_{\,g=g(\,t\,|\,q_+,q_-)}.              \label{2.35}
	\end{eqnarray}
Here $F_{ab}$ is the gauge field operator
	\begin{eqnarray}
	&&F_{ab}=S_{ab}-
	\chi^\mu_a\,c_{\mu\nu}\chi^\nu_b,             \label{2.20}\\
	&&S_{ab}\equiv\frac{\,\delta^2 S\,[\,g\,]}
	{\delta g^a\,\delta g^b},\,\,\,
	\chi^\mu_a\equiv
	\frac{\delta\chi^\mu}{\delta g^a},           \label{2.22}
	\end{eqnarray}
with the functional matrix $\chi^\mu_a$ of linearized gauge conditions --
a first-order differential operator
	$\chi^\mu_a=\chi^\mu_{\;\;a}(d/dt)
	\delta(t-t')$.

The determinants in (\ref{2.35}) are calculated on functional spaces
defined by boundary conditions for gauge and ghost operators. In
relativistic gauge (\ref{2.5}) these operators
$F_{ab}=-a_{ab}\,d^2/dt^2+...$ and
$Q^\mu_\nu=-a^\mu_\nu\,d^2/dt^2+...$ are of second order in time
derivatives with nondegenerate matrices
$a_{ab}=\partial^2 L_{\rm gf}/\partial\dot g^a\partial\dot g^b$ (the Hessian
of gauge-fixed Lagrangian) and (\ref{2.5}). Their boundary conditions were
derived in \cite{barvin} from the conditions (\ref{2.11})-(\ref{2.12}) on
the integration range in the path integral. For $(n+m)\times(n+m)$-matrix
of the gauge field propagator $G^{ab}=G^{ab}(t,t')$ they form
a combined set of $n$ Dirchlet $(i=1,...n)$ plus $m$ Robin $(\mu=1,...m)$
boundary conditions
	\begin{eqnarray}
	&&F_{ca}(d/dt) G^{ab}(t,t')=
	\delta^b_c\delta(t-t'),                \label{2.24}\\
	&&G^{ib}(t_\pm,t')=0,                   \label{2.25}\\
	&&\chi^\mu_{\;\;a}(d/dt_\pm)
	G^{ab}(t_\pm,t')=0                     \label{2.26}
	\end{eqnarray}
and Dirichlet boundary conditions for the ghost propagator
$Q^{-1\,\beta}_{\,\alpha}=Q^{-1\,\beta}_{\,\alpha}(t,t')$
	\begin{eqnarray}
	Q^{\,\;\alpha}_\mu(d/dt)\,Q^{-1\,\beta}_{\,\alpha}(t,t')=
	\delta^\beta_\mu,\,\,\,
	Q^{-1\,\beta}_{\,\alpha}(t_\pm,t')=0.     \label{2.27}
	\end{eqnarray}
Conditions (\ref{2.26}) belong to the Robin type because for
relativistic gauges $\chi^\mu_{\;\,a}(d/dt)$ contains
derivatives transversal to the boundary.

The contribution of the local measure (\ref{2.8}) to
${\mbox{\boldmath$P$}}(q,q')$ in (\ref{2.35}) is
identically rewritten in terms of ${\rm Det}\,a_{ab}$ and
${\rm Det}\,a^\mu_\nu$ -- functional determinants
of ultralocal matrix coefficients of second derivatives in $F_{ab}$ and
$Q^\mu_\nu$. This representation will be used to show the cancellation
of strongest $\delta(0)$-type divergences of ${\rm Det}\,F_{ab}$ and
${\rm Det}\,Q^\mu_\nu$ by the local measure.

Thus, the main goal of this paper consists in the proof of equality of
expressions (\ref{0.12}) and (\ref{2.35}). This proof begins with
the discussion of their gauge independence.

\section{Ward identities and gauge independence}
\hspace{\parindent}
Expressions (\ref{0.12}) and (\ref{2.35}) are independent of the gauge
choice -- the matrices $(X^\mu_i,X^\nu_{k'})$ and $\chi^\mu_a$ respectively.
The mechanism of this gauge independence is based on Ward identities which
we consider in parallel both in canonical and covariant contexts.
The Ward identity for the gauge-fixed Van Vleck matrix (\ref{0.9}) follows
by contracting it with $\nabla^i_\mu$ and using the degeneracy relation
(\ref{0.8})
	\begin{eqnarray}
	C_{\mu\nu}\,X^\nu_{k'}\,D^{-1\;k'i}=
	J^{-1\,\nu}_\mu \nabla^i_\nu.                \label{0.13}
	\end{eqnarray}
As a consequence, arbitrary variations of $(X^\mu_i,X^\nu_{k'},C_{\mu\nu})$
in (\ref{0.12}) vanish due to the cancellation of terms coming from
${\rm det}\,D_{ik'}$ and $JJ'$.

In the covariant (space)time context the analogue of (\ref{0.8}) is the
on-shell degeneracy of the functional Hessian matrix $S_{ab}$ (\ref{2.22})
-- a corollary of (\ref{1.10}),
$R^a_\mu S_{ab}=-\delta R^a_\mu/\delta g^b \delta S/\delta g^a=0$.
It leads to the relation
	$R^a_\mu F_{ab}=-Q^\alpha_\mu
	c_{\alpha\beta} \chi^\beta_b$,
which implies the Ward identity relating the gauge and ghost
propagators subject to boundary conditions of the above type
	\begin{eqnarray}
	c_{\alpha\beta}\stackrel{\rightarrow}{\chi}{\!}^\beta_b
	G^{bc}(t,t')=-Q^{-1\,\beta}_{\,\alpha}(t,t')\!
	\stackrel{\leftarrow}
	{R}{\!}^{\;c}_\beta.       \label{2.37}
	\end{eqnarray}
Here the arrows show the direction in which the derivatives in
differential operators $\chi^\beta_b(d/dt)$ and $R^c_\beta(d/dt')$
are acting on the arguments of Green's functions.

These Ward identities lead to gauge independence of the one-loop
prefactor (\ref{2.35}) provided we consistently fix the definition of
functional determinants ${\rm Det}\,F_{ab}$ and ${\rm Det}\,Q^\mu_\nu$.
Problem is that uniquely defined Green's functions do not yet uniquely
fix these determinants. Variational equations for the latter involve
the functional composition of Green's functions with variations of
their operators: $G^{ba} \delta F_{ab}$ and
$Q^{-1\,\nu}_{\,\mu} \delta Q^\mu_\nu$. However, kernels of propagators
are not smooth functions and their irregularity enhances when they are
acted upon by the derivatives of $\delta F_{ab}(d/dt)$ and
$\delta Q^{\;\;\mu}_\nu(d/dt)$. Therefore, one has to prescribe the way how
these derivatives act on both arguments of Green's functions. We assume
that the varied differential operators $\delta F_{ab}(d/dt)$ and
$\delta Q^\mu_\nu(d/dt)$ are understood as acting in two different ways
	\begin{eqnarray}
	&&\delta\,{\rm ln}\,{\rm Det}\,F_{ab}=
	G^{ba}\!\!\stackrel{\leftrightarrow}
	{\delta F}{\!}_{ab},                            \label{2.39}\\
	&&\delta\,{\rm ln}\,{\rm Det}\,Q^\mu_\nu=
	Q^{-1\,\nu}_{\,\mu}\!\!\stackrel{\leftarrow}
	{\delta Q}{\!}^{\;\;\mu}_\nu.                     \label{2.40}
	\end{eqnarray}
In contrast with the ghost operator, for which both of its derivatives are
acting on {\it one} argument of the Green's function, eq.(\ref{2.39}) here
implies a symmetric action of $\delta F_{ab}(d/dt)$ on {\it both}
arguments of $G^{ba}(t,t')$ in the sense that
	$L^{(2)}_{\rm gf}=(1/2)\varphi^a(t)\!
	\stackrel{\leftrightarrow}
	{F}{\!}_{ab}(d/dt)\,\varphi^b(t)$
represents the quadratic part of the gauge-fixed Lagrangian in perturbations
of field variables $\varphi^a$ \cite{reduct}. It contains the squares of
$\dot{\varphi}^a(t)$ rather than $\ddot{\varphi}^a(t)$ (and
$\stackrel{\leftrightarrow}{\delta F}{\!}_{ab}(d/dt)$ obviously implies
arbitrary variation of this kernel).

With these conventions the gauge variation of gauge and ghost
determinants in (\ref{2.35})
	\begin{eqnarray}
	&&\delta_\chi\,{\rm ln}\,{\rm Det}\,F_{ab}=
	-2\,c_{\alpha\beta}\!\stackrel{\rightarrow}{\chi}{\!}^\beta_b\,
	G^{ba}\,\delta\!\!
	\stackrel{\leftarrow}{\chi}{\!}^\alpha_a,      \label{2.42}\\
	&&\delta_\chi\,{\rm ln}\,{\rm Det}\,Q^\mu_\nu
	=
	Q^{-1\,\beta}_{\,\alpha}\!
	\stackrel{\leftarrow}{R}{\!}^b_\beta\,\delta\!\!
	\stackrel{\leftarrow}{\chi}{\!}^\alpha_b      \label{2.43}
	\end{eqnarray}
cancel out in virtue of (\ref{2.37}). This proves the gauge independence
of (\ref{2.35}) along with fixing the prescription for the
variational definition (\ref{2.39})-(\ref{2.40}) of the functional
determinants\footnote
{
In context of the one-loop effective action Ward identities were
considered in \cite{PhysRep}. Here the main emphasis in their mechanism
is focused on boundary conditions and accurate definition of the
functional determinants. Another choice of these variational
equations leads to extra surface terms violating gauge independence.
}.

\section{Reduction algorithms for functional determinants}
\hspace{\parindent}
Equality of expressions (\ref{0.12}) and (\ref{2.35}) follows from
the reduction algorithms for functional determinants which reduce their
dimensionality from the functional dimensionality of Det's
in (\ref{2.35}) to that of det's in (\ref{0.12}) \cite{reduct}. The simplest
example of such algorithms is the relation (\ref{2.8}) for a purely divergent
local measure. For an ultralocal operator $a_{ab}=a_{ab}(t)\,\delta(t-t')$
one has
	\begin{eqnarray}
	{\rm ln}\,{\rm Det}\,a_{ab}=
	\delta(0)\int_{t_-}^{t_+}dt\,
	{\rm ln}\,{\rm det}\,a_{ab}(t).      \label{2.10}
	\end{eqnarray}
For differential operators the reduction algorithms result from the
functional integration of eqs.(\ref{2.39})-(\ref{2.40}). For symmetric
operators with Dirichlet boundary conditions they were considered in much
detail in \cite{reduct}. Here we generalize them to the case of combined
Dirichlet-Robin boundary conditions (\ref{2.25})-(\ref{2.26}) for $F_{ab}$
and to the case of non-symmetric ghost operator $Q^\mu_\nu$.

Starting with $F_{ab}(d/dt)$, we note that this operator
satisfies the Wronskian relation with arbitrary test functions
$\varphi^a_{1,2}$
        \begin{equation}
	\varphi^a\,(F_{ab}\varphi^b)-
	(F_{ba}\varphi^a)\,\varphi^b=
	-\frac{d}{dt}\left[\,\varphi^a\,
	(W_{ab}\varphi^b)-
	(W_{ba}\varphi^a)\,\varphi^b\,\right],	  \label{3.9}
	\end{equation}
where $W_{ab}=a_{ab}\,d/dt+...$ is the {\it Wronskian}
operator. It participates in the variational equation for the canonical
momentum,
	  $\delta(\partial L_{\rm gf}/\partial\dot g)
          =W(d/dt)\,\delta g\,(t)$,
and, in particular, has a special form of its $\mu$-component given by the
operator of linearized gauge conditions
	$W_{\mu b}(d/dt)=a^\alpha_\mu c_{\alpha\beta}\chi^\beta_b(d/dt)$.

Similarly to \cite{reduct}, introduce for $F_{ab}(d/dt)$ two complete sets of
basis functions $u_-^a(t)$ and $u_+^a(t)$
        \begin{eqnarray}
	&&F_{ab}(d/dt)\,u_{\pm}^b(t)=0,  \label{3.17}\\
	&&u_+^i(t_+)=0,\,\,
	\stackrel{\rightarrow}{W}_{\mu a}\!
	u_+^a(t_+)=0,                             \label{3.18}\\
	&&u_-^i(t_-)=0,\,\,
	\stackrel{\rightarrow}{W}_{\mu a}\!
	u_-^a(t_-)=0.                              \label{3.19}
	\end{eqnarray}
satisfying the boundary conditions (\ref{2.25})-(\ref{2.26})
respectively at $t_{-}$ and $t_{+}$, the Robin conditions
(\ref{2.26}) being rewritten in terms of $W_{\mu b}(d/dt)$. We assume that
the indices enumerating complete sets of these basis
functions are encoded in subscripts $\pm$. Then, the $t$-independent
Wronskian inner product of basis functions forms a matrix in vector space of
these indices. In view of boundary conditions it has only
two nonvanishing blocks given by two mutually transposed matrices
	\begin{eqnarray}
	{\mbox{\boldmath$\Delta$}}_{-+}=
	u^a_{-}\,(Wu_{+}\!)_a-(Wu_{-}\!)_a\,u_{+}^b,
	\,\,\,{\mbox{\boldmath$\Delta$}}_{+-}=
	-{\mbox{\boldmath$\Delta$}}_{-+}^T,            \label{3.22}
	  \end{eqnarray}
where for brevity we introduced the notation
	$(Wu_\pm)_a\equiv\,
	\stackrel{\rightarrow}{W}_{ab}\!u^b_\pm$.
In terms of basis functions the Green's function of the mixed
Dirichlet-Robin boundary value problem (\ref{2.24})-(\ref{2.26}) has the
following representation
	\begin{eqnarray}
	&&G^{ab}(t,t')=
	-\theta\,(t\!-\!t')\,u_{\!+}^a(t)\,
	({\mbox{\boldmath$\Delta$}}_{-+}\!)^{-1}
	u_-^b(t')
	\nonumber\\
	&&\qquad\qquad\qquad\qquad\qquad+\theta\,(t'\!-\!t)\,
	   u_{\!-}^a(t)\,
	({\mbox{\boldmath$\Delta$}}_{+-}\!)^{-1}
	u_{+}^b(t'),       \label{3.26}
	\end{eqnarray}
where $\theta(x)$ is the step function: $\theta(x)=1$ for $x>0$
and $\theta(x)=0$ for $x<0$.

After substituting (\ref{3.26}) to (\ref{2.39}) the $\delta(0)$-type
terms, which originate from the derivatives of $\theta$-functions,
reduce to the variation of (\ref{2.10}) \cite{reduct},
so that the strongest divergences of ${\rm Det}\,F_{ab}$ are cancelled by
the local measure \cite{Bern}. The rest can be transformed by
integrating by parts and using the equations (\ref{3.17}) or
their corollary $F_{ab}u^b_\pm=-F_{ab}\,\delta u^b_\pm$. This
allows one in a systematic way to reduce the full answer to surface terms
at $t=t_\pm$ which form the total variation that can easily be
functionally integrated to give (see \cite{reduc1} for details)
       \begin{eqnarray}
	\left(\,\frac{{\rm Det}\!\stackrel{\leftrightarrow}{F}_{\!ab}}
	{{\rm Det}\,
	a_{ab}}\,\right)^{\!-1/2}=
	{\rm const}\,\left(\,\frac{{\rm det}\,
	\left(\,{\mbox{\boldmath$S$}}_{ik'}
	+X^\mu_iC_{\mu\nu}X^\nu_{k'}\,\right) }
	{{\rm det}\,C_{\mu\nu}}\,\right)^{\!1/2},     \label{3.41}
	\end{eqnarray}
where the matrices $(X^\mu_i,X^\nu_{k'},C_{\mu\nu})$ in terms of
basis functions of the above type read
	\begin{eqnarray}
	&&X^\nu_i=(Wu_+)_i(t_+)
	\,{\mbox{\boldmath$\Delta$}}_{-+}^{-1}u^\nu_-(t_-), \label{3.35}\\
	&&X^\mu_{k'}=u^\mu_+(t_+)\,
	{\mbox{\boldmath$\Delta$}}_{-+}^{-1}\,
	(Wu_-)_k(t_-),                              \label{3.36}\\
	&&C^{\mu\nu}=u^\mu_+(t_+)\,
	{\mbox{\boldmath$\Delta$}}_{-+}^{-1}\,
	u^\nu_-(t_-),\,\,\,C_{\mu\nu}=(C^{\mu\nu})^{-1}   \label{3.37}
	\end{eqnarray}
and ${\mbox{\boldmath$S$}}_{ik'}$ arising here as
	${\mbox{\boldmath$S$}}_{ik'}=-(Wu_+)_i(t_+)
	\,{\mbox{\boldmath$\Delta$}}_{-+}^{-1}\,(Wu_-)_k(t_-)$
coincides with the Van Vleck matrix (\ref{0.6}) for the principal Hamilton
function \cite{reduc1}. This equation is a generalization of the results
of paper \cite{reduct} extending the path-integral derivation of the
Pauli-Van Vleck-Morette formula \cite{Morette} to gauge theories. Its
comparison with
(\ref{0.9}),(\ref{0.12}) shows that the quantities (\ref{3.35})-(\ref{3.37})
are special unitary gauge conditions and gauge-fixing matrix induced by the
relativistic gauge fixing procedure.

The functional integration of eq.(\ref{2.40}) for
${\rm Det}\!\stackrel{\rightarrow}{Q}{\!}^\mu_\nu$ repeates the calculations
of \cite{reduct} with modifications caused by the asymmetry of the ghost
operator. This asymmetry results in a double set of right and left basis
functions which give rise to the representation of $Q^{-1\,\beta}_{\,\alpha}$
similar to (\ref{3.26})
and after functional integration yield the analogue of the Pauli-Van Vleck
formula with a special Van Vleck matrix \cite{reduc1}. In view of Ward
identity (\ref{2.37}) special combinations of {\it ghost} field basis
functions entering this matrix can be related to the
unitary gauge conditions (\ref{3.35})-(\ref{3.36}) built in terms of the
basis functions of the {\it gauge} operator. The final reduction algorithm
then reads (see \cite{reduc1} for details)
	\begin{eqnarray}
	&&\frac{{\rm Det}\!\stackrel{\rightarrow}{Q}{\!}^\mu_\nu}
	{{\rm Det}\,a^\mu_\nu}=
	{\rm const}\,\big(\,{\rm det}\,J^\mu_\nu\,
	{\rm det}\,J'^\mu_\nu\,\big)^{-1/2},            \label{4.24}
	\end{eqnarray}
where the matrices $J^\mu_\nu$ and $J'^\mu_\nu$ coincide with unitary
Faddeev-Popov operators (\ref{0.10})-(\ref{0.11}) in unitary gauges
of the above type.

The substitution of (\ref{3.41}) and (\ref{4.24}) to (\ref{2.35})
accomplishes the proof of the needed equality of (\ref{0.12}) and
(\ref{2.35}).

\section{Conclusions}
\hspace{\parindent}
The virtue of equivalence of (\ref{0.12}) and (\ref{2.35})
is that it establishes the explicit mechanism of transition from
relativistic to unitary gauge conditions. The relations between
them are nonlocal in time -- the matrices of unitary gauge-fixing
procedure (\ref{3.35})-(\ref{3.37}) express in terms of basis functions
of the gauge field operator, nonlocally depending on its relativistic gauge.
Such a transition is very important because it proves intrinsic unitarity of
a manifestly covariant quantization in terms of a Lagrangian path
integral, and the mechanism of this transition revealed here is
free from singularities inherent to the usual $\epsilon$-procedure of
formally identical transformations in the path integral \cite{FV}.

In physical applications, Feynman diagrammatic expansion of the path
integral as a means of solving {\it noncovariant} quantum Dirac constraints
\cite{BKief} is of crucial importance due to {\it spacetime covariance} of
their solution that can be attained by a suitable choice of relativistic
gauge conditions. Important implications of this technique belong to quantum
cosmology of the early universe \cite{qcr1} where correct predictions can
be achieved only within spacetime covariant approach to loop effects.

The aspects of gauge independence considered above are important
for problems in spacetimes with boundaries or nontrivial
time foliations. There exists a long list of gauge-dependent results for
a formally gauge independent quantity -- the one-loop effective action
\cite{GrifKos}. Exhaustive explanation of these discrepancies
can be expected on the basis of Ward identities with a special
emphasis on boundary conditions.

Finally, the physics of wormholes in Euclidean quantum gravity
\cite{worm} also belongs to the scope of our result. The predictions
of this theory are based, in particular, on
the existence of a negative mode on the wormhole instanton \cite{RubShved} --
a formal extrapolation of the mechanism applicable only to
non-gravitational systems \cite{Coleman}. Thus, these predictions should
be revised from the viewpoint of the Wheeler-DeWitt equation
\cite{TanSas,RubShved} -- the gravitational quantum Dirac constraint.
This negative mode belongs to the {\it nondynamical} conformal sector,
and in the Lorentzian theory its contribution is cancelled by ghost fields in
relativistic gauges. Therefore, one should expect a similar cancellation
by the mechanism of Ward identities also in Euclidean theory. Other
problems related to the idefiniteness of the Euclidean gravitational
action include the lack of strong ellipticity of the Dirichlet-Robin
boundary value problem (\ref{2.24})-(\ref{2.26}) \cite{avesp1}. The
proposed technique is a direct avenue towards the
resolution of these issues which are currently under study.

\section*{Acknowledgements}
\hspace{\parindent}
This work was supported by the Russian Foundation for Basic Research
under grants 96-02-16287 and 96-02-16295 and the European
Community Grant INTAS-93-493-ext. Partly this work has been made possible
also due to the support by the Russian Research Project
``Cosmomicrophysics''.

\end{document}